# Microstructure and thermal properties of unalloyed tungsten deposited by Wire + Arc Additive Manufacturing

G. Marinelli[a,*], F. Martina[a], S. Ganguly[a], S. Williams[a], H. Lewtas[b], D. Hancock[b], S. Mehraban[c], N. Lavery[c]

[a]Welding Engineering and Laser Processing Centre (WELPC), Cranfield University, Cranfield, MK43 0AL, United Kingdom

[b]Culham Centre for Fusion Energy (CCFE), Culham Science Centre, Abingdon, OX14 3DB, United Kingdom

[c]Department of Materials Engineering, Singleton Abbey, Singleton Park Campus, Swansea, SA2 8PP, United Kingdom

[*]Corresponding Author. E-mail address: g.marinelli@cranfield.ac.uk (G. Marinelli)

**Abstract**

Tungsten is considered as one of the most promising materials for nuclear fusion reactor chamber applications. Wire + Arc Additive Manufacturing has already demonstrated the ability to deposit defect-free large-scale tungsten structures, with considerable deposition rates. In this study, the microstructure of the as-deposited and heat-treated material has been characterised; it featured mainly large elongated grains for both conditions. The heat treatment at 1273 K for 6 hours had a negligible effect on microstructure and on thermal diffusivity. Furthermore, the linear coefficient of thermal expansion was in the range of $4.5 \times 10^{-6}$ µm m$^{-1}$ K$^{-1}$ to $6.8 \times 10^{-6}$ µm m$^{-1}$ K$^{-1}$; the density of the deposit was as high as 99.4% of the theoretical tungsten density; the thermal diffusivity and the thermal conductivity were measured and calculated, respectively, and seen to decrease considerably in the temperature range between 300 K to 1300 K, for both testing conditions. These results showed that Wire + Arc Additive Manufacturing can be considered as a suitable technology for the production of tungsten components for the nuclear sector.

## 1 Introduction

The development of innovative sources of energy has always been accompanied by improvements in production techniques and material properties, needed to ensure optimal energy conversion rates [1]. The controlled extraction of energy from the nuclear fusion reaction is one of the major challenges today's energy sector is facing [2]. Considered such harsh operating environment, tungsten has been considered as the major candidate for plasma facing material for several years [3–5]. For some components' concepts f.i. the helium-cooled divertor and helium-cooled first wall, it has also been considered for partially-structural roles [2,4,6], given its excellent mechanical properties at a wide range of temperature, adequate heat resistance, and high neutron load capacity [2,7]. Tungsten's adoption is also aided by its high melting point, high density, relatively high thermal conductivity at room temperature, and low coefficient of thermal expansion [8]. Indeed, heat flux tests have concluded that the maximum operating temperature of the helium-cooled divertor should be 800 K for the blanket first-wall; around 1000 K – 1200 K for the lowest armour; and above 2000 K for the armour surface [2], making tungsten an ideal candidate material. Finally, tungsten is characterised by a high resistance to sputtering or erosion, and low tritium retention [9,10].

Currently, tungsten components are almost exclusively manufactured either via the powder-metallurgy route or via hot-forming processes [11]; indeed casting is impractical due to tungsten's extremely high melting point. The additions of sintering and cold/hot rolling enable fine control over the final microstructures [11]. However, some undesirable properties such as the high ductile-to-brittle-transition temperature (DBTT) and the high recrystallization temperature still limit the adoption of tungsten and increase components' cost. Finally, as machining is necessary to produce complex geometries [2], applications are limited further due to tungsten's hardness and brittleness.

A solution to these issues could be brought about by Additive Manufacturing (AM) [12,13]. The layer-by-layer approach can enable substantial cost reduction, freedom of design and control over mechanical properties [14,15]. For a comprehensive review of the potential of AM of metals, please refer to the work of Frazier [16] and Herderick [17]. Within the range of possible AM technologies, wire-feed processes are very suitable for the production of large-scale components [18]. In particular, Wire + Arc additive Manufacturing (WAAM), which employs an electric arc as the heat source, and high-quality metal wire as the feedstock [19], can directly fabricate fully-dense large 3-D near-net-shape components, at much higher rates, than most other metal AM processes [18,19], the highest rate so far being of 9.5 kg/h [20]. The WAAM process has successfully produced large-scale parts in stainless steel [21], Inconel ® [22], titanium [23] and aluminium [24]. Furthermore, a previous research showed that unalloyed tungsten can be deposited via WAAM with a noteworthy absence of micro-cracks among the layers [25].

The manufacture of large and engineered components by WAAM is attractive also because of the low system and operating costs, as well as the modularity of the system design [19,26].

Therefore, the deposition of tungsten by WAAM is of interest to the nuclear fusion community. The requirements for a tungsten structural component for fusion environment, made by WAAM, would include high thermal conductivity, high-temperature stability, high recrystallization temperature, and sufficient ductility to operate under constant neutron irradiation [2]. Furthermore, isotropy of the microstructure would be desirable, as the alignment of the grains could influence the behaviour of the tungsten tiles under high-heat fluxes [2]. Control of these microstructural features can possibly lead to a more uniform stress distribution, the best operating condition to minimise the risk of failures [2].

To the authors' best knowledge, none of the publications on AM of tungsten has discussed its thermal properties; moreover, WAAM of tungsten has been only reported in our previous research [25]. In this study, the microstructure and the thermal properties of unalloyed tungsten deposited by WAAM have been characterised in both the as-deposited and heat-treated conditions. In particular, the microstructures along all three principal directions have been analysed. The coefficient of linear thermal expansion, the thermal diffusivity and the heat capacity have been measured over a range of temperatures using dilatometry and laser flash analysis. The thermal conductivity was calculated from diffusivity, heat capacity and density. The main aim of this study was to prove that the WAAM process is indeed capable to manufacture large-scale unalloyed tungsten components, with levels of structural integrity suitable for nuclear applications.

## 2    Material and methods

Unalloyed tungsten wire (1 mm diameter) was used as feedstock. An unalloyed tungsten plate (210 mm in length, 10 mm in width and 30 mm in height) produced by powder metallurgy was used as substrate, needed to start the deposition. The surfaces of the plate were ground and rinsed with acetone prior to starting, to remove most surface contaminants.

**Fig. 1** shows the layout of the WAAM apparatus. A conventional tungsten inert gas (TIG) welding torch, a power supply and a controlled wire feeder were used. The heat source, the wire delivery system and the substrate were attached to three motorized linear stages in the XYZ configuration. The substrate was firmly held by a clamping system; in particular, a lateral force was applied to the substrate by two copper bars, using a series of 5 screws per side. The wire was fed from the front of the weld pool [25], and the direction of deposition was always the same. The width of the deposit was that of a single deposited bead (12 mm), given by the parameters shown in **Table 1**; these were kept constant throughout the entire deposition. The average height of a layer was around 1.2 mm. Please note the choice of 100% He as process-gas for TIG, following from what concluded in our previous study [27]. The total height and length of the deposit were 75 mm and 120 mm, respectively. The entire apparatus was contained within an air-tight enclosure, to achieve an $O_2$ concentration of around 100 ppm.

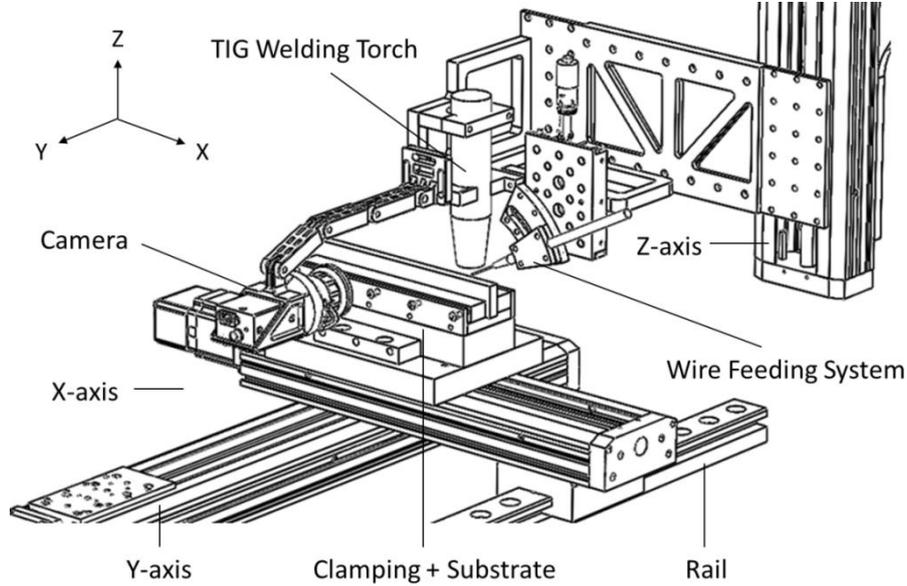

**Fig. 1**: Set-up used for the deposition of the tungsten structure via WAAM.

**Table 1**
Wire + Arc Additive Manufacturing process parameters used for the deposition of unalloyed tungsten.

| | |
|---|---|
| **Travel speed (TS) [mm/s]** | 2 |
| **Welding Current (I) [A]** | 400 |
| **Wire Feed Speed (WFS) [mm/s]** | 35 |
| **Electrode-workpiece Distance [mm]** | 3.5 |
| **Electrode angle [°]** | 45 |
| **Electrode Diameter [mm]** | 3.2 |
| **Shielding gas composition [%]** | 100 He |
| **Gas flow rate [L/min]** | 15 |

X-ray fluorescence (XRF) spectroscopy was used for the detection of the major metallic atoms (W, Mo, Ta, Ti, V, Cr and Fe). LECO combustion analysis was used for the measurement of the concentration of carbon, nitrogen and oxygen. Finally, inductively coupled plasma optical emission spectroscopy (ICP-OES) was used for the detection of potassium.

12 samples were extracted for laser flash analysis (LFA), their dimensions were 10 mm in height, 10 mm in width and 3 mm in thickness. 12 cylindrical samples were extracted for dilatometry (DIL), with a total length of 20 mm and a diameter of 5 mm (**Fig. 2a**). Finally, six samples were extracted for microstructural analysis. The extraction of the samples was conducted using electrical discharge machining (EDM). Half of the coupons were tested in the as-deposited condition ("As-Dep"), while the other half was tested after heat treatment at 1273 K for 6 hours under vacuum with a heating rate of 5 K per minute ("Heat-Treat").

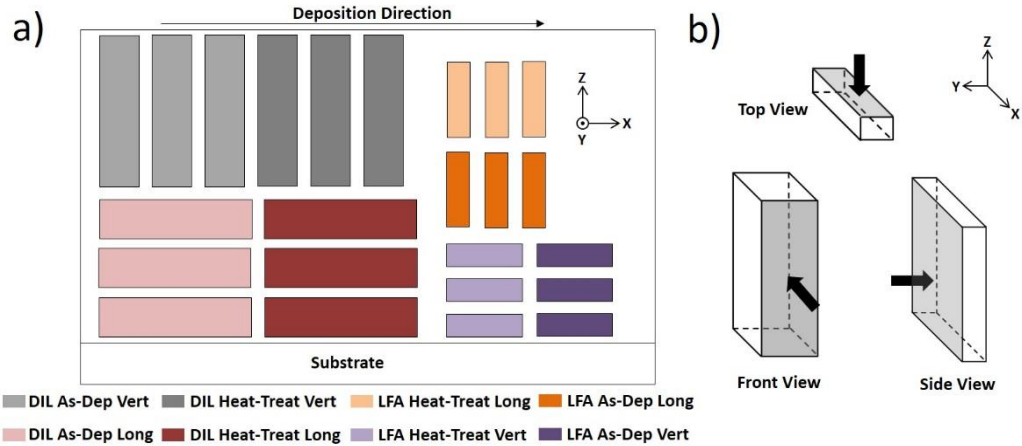

**Fig. 2**: Extraction plan of the coupon with respect to the deposition direction with the denomination of the coupons (a); Plane used for microstructural analysis and their denomination (b).

The microstructure of the component in the As-Dep and Heat-Treat condition was examined using three different cross-sections; the nomenclature is clarified in **Fig. 2b**; "Front View", "Side View" and "Top View" show the microstructural features of the YZ, XZ and XY planes, respectively.

A Netzsch DIL 402 was used for the dilatometry. The initial length of each sample was measured using a micrometre; each sample was heated from 293 K to 1473 K at a rate of 10 K/min under an argon atmosphere, and then cooled down to 293 K at a rate of 10 K/min. The variations in length during expansion (dL) and contraction were recorded using the Netzsch Proteus software. A known alumina standard was used as a correction sample to remove the effective thermal expansion of the sample carrier. The true coefficient of linear thermal expansion ($\alpha_t$) was calculated by using the derivative dL/dT at a single temperature according to **Eq. (1)**:

$$\alpha_t(T) = \frac{1}{L_0}\left(\frac{\partial L}{\partial T}\right)_p \tag{1}$$

Where $L_0$ is the length at 293 K, T is the temperature at the time of measurement, and L is the specimens' length at the time of measurement.

The thermal diffusivity of each sample was characterised using a Netzsch LFA 457. The thickness of each sample was firstly measured using a micrometre; then the samples were coated with a layer of graphite, to increase the absorption of the laser radiation and to reduce reflection. Each sample was placed in a sample holder made of aluminium titanate with a silicon carbide cap. In particular, a three-position sample holder was used and it was filled with two tungsten samples and a Pyroceram 9606 for standard reference. The sample holder was inserted in the high-temperature furnace chamber which was evacuated and backfilled with argon. A sustained flow of 100 ml per minute of argon was used throughout the experiment. The sample's temperature change was detected using an InSn IR detector cooled by liquid nitrogen. Temperature steps of 50 K were taken within the 293 K to 1273 K temperature range. Five shots were performed at each temperature step and the individual diffusivity results measured and then averaged. The thermal diffusivity of each shot was calculated using the Cowon method with pulse correction applied [28]. This model takes into account heat loss from all surfaces of the sample. The values of thermal diffusivity for different temperature were calculated using the Netzsch Proteus software. From the thermal diffusivity data, the specific heat capacity was also calculated using the Pyroceram reference and the ratio method (**Eq. (2)**):

$$C_p^{Sample} = \frac{T_\infty^{Ref.}}{T_\infty^{Sample}} \cdot \frac{Q^{Sample}}{Q^{Ref.}} \cdot \frac{V^{Sample}}{V^{Ref.}} \cdot \frac{\rho^{Ref.} \cdot D^{Ref.}}{\rho^{Sample} \cdot D^{Sample}} \cdot \frac{d^{2,sample\ Orifice}}{d^{2,Ref.Orifice}} \cdot C_p^{Ref.}(T) \qquad (2)$$

The bulk density of each sample was measured using the method of hydrostatic weighing, employing Archimedes' principle [29]. A tensiometer was used to determine the weight of each sample in air and in distilled water. From the difference between these two values (ΔW), the sample density can be calculated using the **Eq. (3)**:

$$\rho_{sample} = Z * \frac{W}{\Delta W} \qquad (3)$$

Where $\rho_{sample}$ is the sample density, Z is the water density, W is the sample weight in air, and ΔW is the difference between sample weight in air and samples weight in water. The value of the water density at 300 K was used for calculation. The change in density during heating was also considered by using the thermal expansion data from the DIL measurements. The temperature-dependent density was calculated using the **Eq. (4)**:

$$\rho(T) = \frac{\rho_0}{\left(1 + \frac{\Delta L}{L_0}(T)\right)^3} \qquad (4)$$

Where $\rho(T)$ is the density at a specific temperature, $\rho_0$ is the measured density, and $\Delta L/L_0$ is the variation in length measured by dilatometry. The thermal conductivity was calculated from the thermal diffusivity, the specific heat and the density as measured and calculated, according to the Laplace relation (**Eq. (5)**):

$$k(T) = \alpha(T) \cdot \rho(T) \cdot C_p(T) \qquad (5)$$

Where k is the thermal conductivity, α is the thermal diffusivity, ρ is the density, and $C_p$ is the specific heat capacity.

## 3    Results and Discussion

### 3.1.    Appearance and chemical analysis

**Fig. 3** shows the tungsten structure built for this study. The layers were regular and smooth. Despite tungsten's affinity to oxidation at high temperature, the inert argon atmosphere was effective, as can be assessed visually by the shiny silver appearance of the deposit.

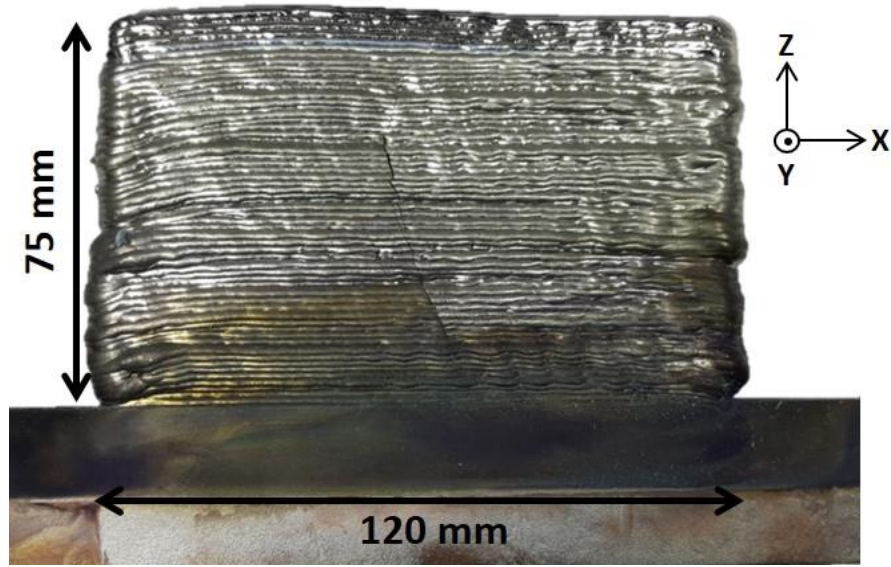

**Fig. 3**: Large-scale unalloyed tungsten linear structure deposited via WAAM process.

A comparison between the chemical composition of wire feedstock and that of the bulk of the deposited material is reported in **Table 2**. It can be concluded that there was no contamination during the deposition process.

**Table 2**
Elemental composition (wt.%) of tungsten substrate, wire and wall used in this study.

|  | W | Mo | Ta | Ti | V | Cr | Fe | C | N | O | K |
|---|---|---|---|---|---|---|---|---|---|---|---|
| **Substrate** | 99.99 | <0.05 | <0.05 | <0.05 | <0.05 | <0.05 | <0.05 | <10 ppm | <10 ppm | <50 ppm | <10 ppm |
| **Wire** | 99.99 | <0.05 | <0.05 | <0.05 | <0.05 | <0.05 | <0.05 | 33 ppm | <10 ppm | <50 ppm | <10 ppm |
| **Deposit** | 99.99 | <0.05 | <0.05 | <0.05 | <0.05 | <0.05 | <0.05 | <10 ppm | <10 ppm | <50 ppm | <10 ppm |

*3.2. Microstructure*

**Fig. 4** reports the bulk microstructure of both As-Dep and Heat-Treat structure in the YZ plane (Front View). There was a marked symmetry of grain shape and orientation with respect to the centreline of the deposit. In particular, two specular arrays of columnar grains inclined at around 45°, both pointing toward the centre of the structure, formed during solidification. The presence of these grains was more predominant on the upper part of both samples. Some clusters of finer equiaxed grains were also found towards the lower part of the samples.

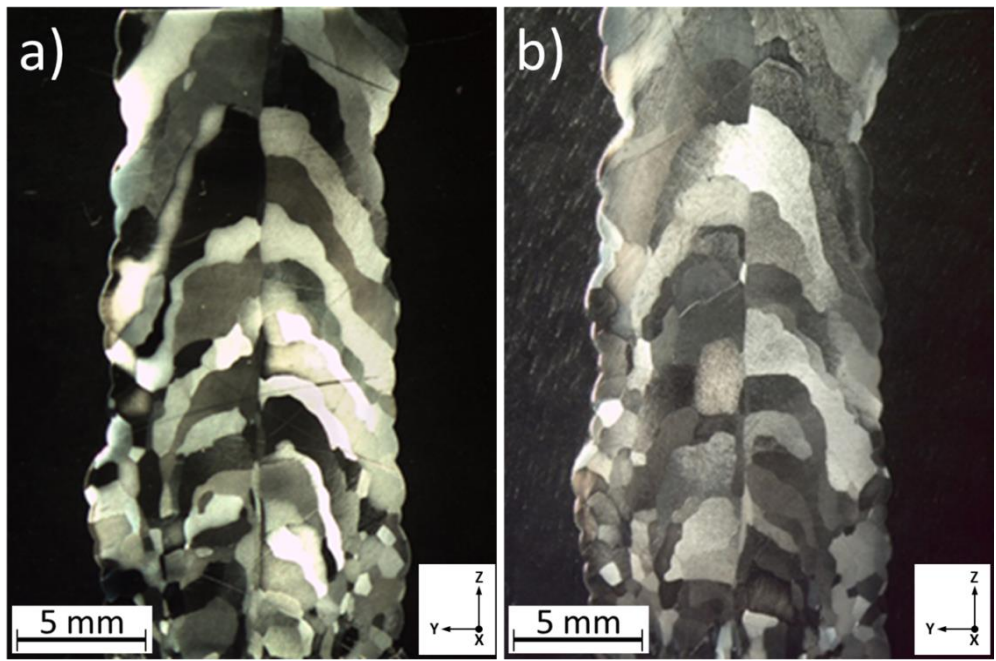

**Fig. 4**: Microstructure of the Front View of the As-Dep (a) and the Heat-Treat structure (b).

The heat treatment process did not have any significant effect on the microstructure: the shape, size and orientation of the grains are similar for both cross-sections. In this study, the sample was heat-treated at a temperature lower than the recrystallisation's one typically reported for tungsten (above 1500 K [30]). Therefore, neither crystal reorientations nor grain boundary migrations occurred during the treatment.

Optical microscopy images of the other two sectioning planes are reported in **Fig. 5** for the As-Dep case only. The final 27 mm of the layer located at Z = 60 mm from the substrate are shown in **Fig. 5a**; two specular arrays of grains can be seen here too. These grains were part of a regular pattern in which each elongated grain grew from the side toward the centre of the layers. Furthermore, each grain seemed to be slightly inclined with respect to the deposition direction. Toward the very end of the deposit, the elongated grain progressively changed their orientation and the presence of few finer grains was also observed; this was due to the faster solidification and cooling conditions seen at the extremity of the deposit. **Fig. 5b** reports the microstructure of the Side View, in which a consistent array of long and parallel columnar grains can be observed.

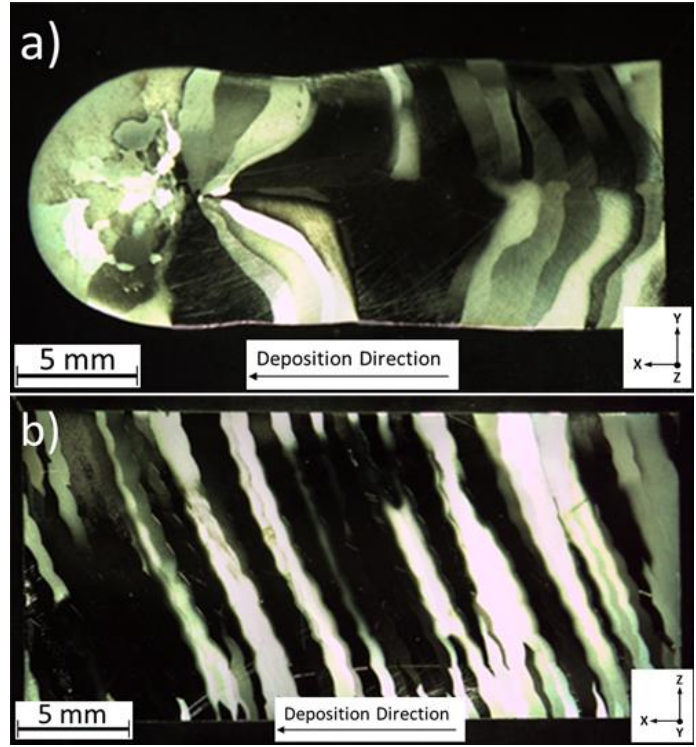

**Fig. 5**: Microstructure of the As-Dep sample, as seen from (a) Top View of the final 27 mm of the layer located at Z = 60 mm from the substrate, and (b) Side View of the deposit's centreline.

In particular, each grain was also inclined at approximately 45°, pointing evidently towards the direction of deposition. This is corroborated by the lateral view of the melt pool reported in **Fig. 6**. The back of the melt pool is inclined due to the moving heat source [31].

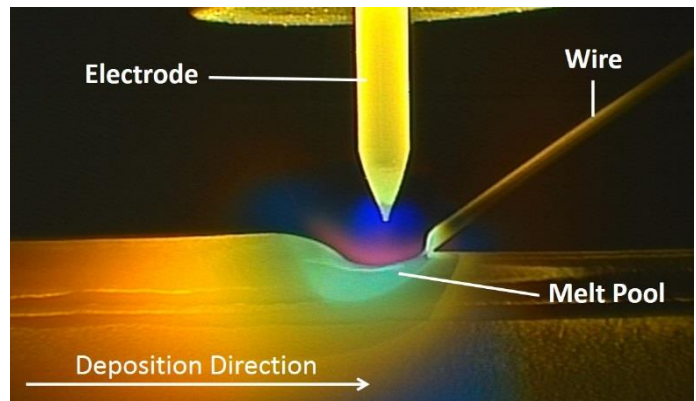

**Fig. 6**: Lateral image of the deposition process.

In summary, the microstructure of single-bead unalloyed tungsten produced by WAAM can be described as composed of two main series of large columnar grains, related by a specular symmetry with respect to the structure's centreline.

**Fig. 7** shows high-magnification optical microscopy images of both Top View (**Fig. 7a**) and Side View (**Fig. 7b**). A few localised pores were found only around the very end of the deposit, where finer grains can also be seen. Furthermore, wavy grain edges were observed (**Fig. 7b**), possibly caused by the re-melting associated with the repetitive nature of the WAAM process.

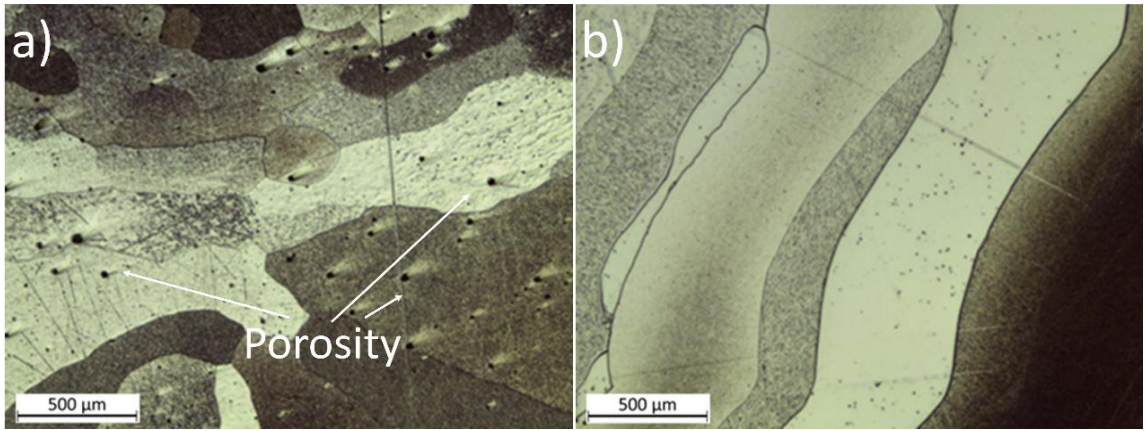

**Fig. 7**: Microstructural detail of the As-Dep structure from the Top View showing porosity at the end of the deposition (a) and from the Side View showing wavy grain boundaries (b).

The grains at the interface between the deposit and the substrate are smaller than that in the bulk of the deposit, and appear also equiaxed (bottom of the images in **Fig. 4**). This is due to the substrate acting as heat-sink, the relatively high thermal conductivity of tungsten around room temperature (the substrate was not preheated prior to the deposition), and the high thermal gradient in that region.

After a few layers, the repetitive heat cycles associated with several layers result in a build-up of temperature and therefore in reduced thermal gradient. This because the thermal conductivity of tungsten decreases as the temperature increases (see Section 3.7 and [32–34]). Consequently, the front of solidification advances parallel to the main heat flow direction. Columnar grains grow epitaxially from the partially-melted grains, at the bottom of the weld pool (**Fig. 4**).

Towards the top of the deposit, the heat accumulation was severe. At very high temperatures, radiation plays a considerable role in the dissipation of heat. In particular, the effectiveness of the radiation heat transfer increases as the temperature increases (according to the Stefan-Boltzmann Law); moreover, tungsten is characterised by a relatively high emissivity [35]. During the deposition, the material around the weld pool eventually reaches a high enough temperature for the radiation to affect the heat flow significantly. Thus, higher-thermal-conductivity regions are established locally on the sides of the weld pool, which are cooled slightly more than the bulk of the material below the weld pool. For this reason, additional lateral heat flows compete with the main flow through the bulk causing the heat to be extracted slightly more efficiently toward the sides of the structure. As the front of solidification usually advances following the direction of the heat extraction, the columnar grains start to solidify with a different angle with respect to lower grains.

*3.3. Thermal Expansion*

The typical thermal expansion curves from 300 K to 1473 K for the As-Dep and Heat-Treat samples are reported in **Fig. 8a** and **Fig. 8b**, respectively. Regardless of the treatment, the material expanded with almost linear dependence on the temperature. Substantial differences can be seen with regards to the testing direction in the As-Dep conditions. As presented in **Fig. 8a,** the samples extracted in the vertical direction (As-Dep Vert) showed a narrow hysteresis amplitude whereas the samples extracted in the longitudinal direction (As-Dep Long) had a much wider hysteresis amplitude. Furthermore, the heating curve of the As-Dep Long sample recorded a change in slope at around 1050 K, which led to a residual strain of 4.55 mStrain at around 450 K, after cooling. No sudden changes were seen in the slope of the thermal expansion curves of the As-Dep Vert samples. The Heat-Treat specimens showed consistent hysteresis width regardless of the testing direction; maximum levels of strain were seen at 1470 K, and they were similar to those of the As-Dep Vert sample.

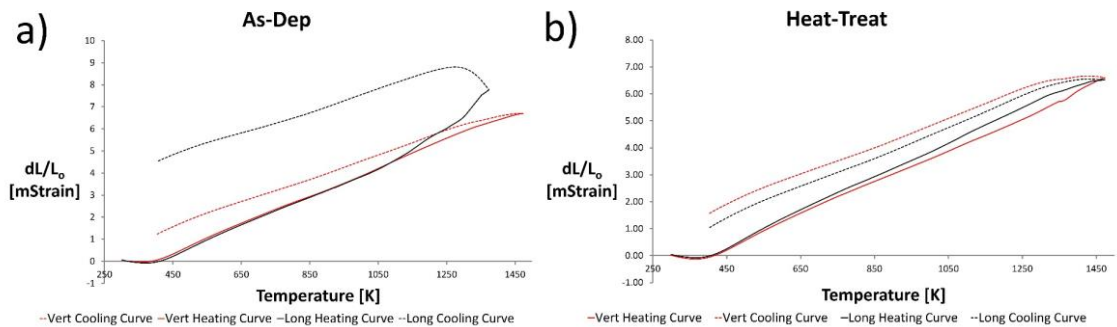

**Fig. 8**: Typical thermal expansion curves of the tested samples in the As-Dep (a) and Heat-Treat condition (b).

The variation of the true coefficient of thermal expansion ($\alpha_t$) between 300 K and 1473 K is reported in **Fig. 9** for the As-Dep and Heat-Treat samples. All samples were characterised by a negative $\alpha_t$ at the beginning of the test. This is due to the initial heating stage within the instrument where the sample carrier is heated before the sample, due to the radial furnace. Then, $\alpha_t$ started to stabilise at around 450 K. Between 1050 K and 1473 K, pronounced variations were recorded, especially for the As-Dep Long samples. On average, the linear coefficient of thermal expansion resulted to be between $4.5 \times 10^{-6}$ μm m$^{-1}$ K$^{-1}$ and $6.8 \times 10^{-6}$ μm m$^{-1}$ K$^{-1}$ for both As-Dep and Heat-Treat samples.

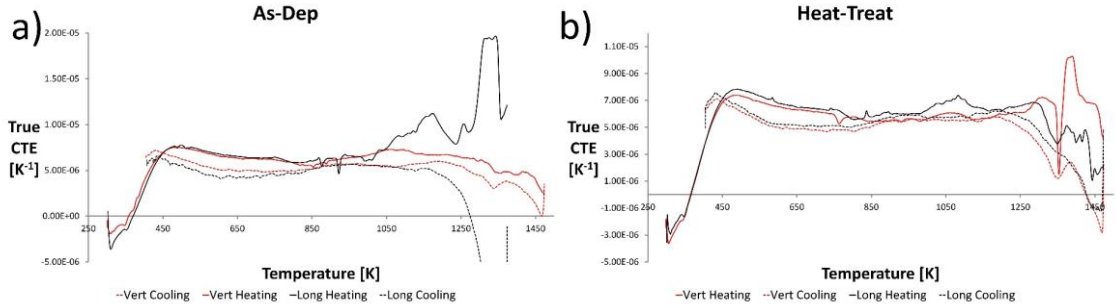

**Fig. 9**: The variation of the $\alpha_t$ over temperature for the heating and cooling phase of the samples in the As-Dep (a) and Heat-Treat condition (b).

The higher maximum strain on heating, and the higher residual strain on cooling of the As-Dep Long samples (if compared to the As-Dep Vert samples) could be caused by the anisotropy of the microstructure (orientation and density of the grain boundaries), and by the presence of residual stress, which are typical of AM structures [36]. Indeed, Type II residual stresses remain also after specimens extraction, as they are related to intergranular stresses [37].

The absence of abrupt changes in strain for the Heat-Treat samples has been attributed to residual stress relaxation. Additionally, the heat treatment could have promoted the dissolution of interstitial from the grain boundaries to the tungsten bulk.

### 3.4. Thermal Diffusivity

Thermal diffusivity is plotted in **Fig. 10**. In general, it decreases of around 44% as the temperature increases from 300 K to 1300 K for all the cases analysed. Thus, the thermal diffusivity was not neither influenced by the direction of testing nor the testing conditions.

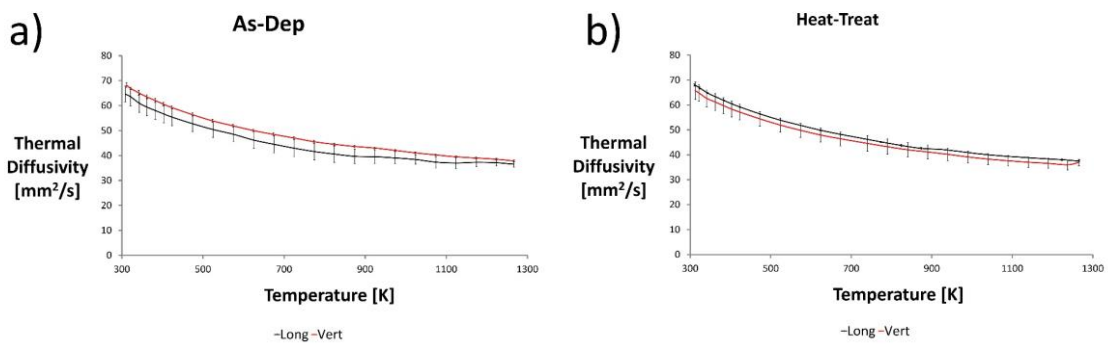

**Fig. 10**: Thermal diffusivity curves against temperate of the tested samples in the As-Dep (a) and Heat-Treat condition (b) with the associated error bars.

*3.5. Density*

The theoretical density of tungsten at 300 K is 19.3 g/cm$^3$ [38]. **Table 3** reports the measured values of the average density ($\rho$) and the values of calculated density at 800 K and 1300 K using **Eq. 4**, which consider the variation of density caused by the expansion of the material. The density in percentage with respect to the theoretical value ($\rho_R$) is also reported in **Table 3**. The As-Dep samples had a lower average density when compared to the Heat-Treat samples, which reaches 99.4 % of the theoretical density. This modest difference between As-Dep and Heat-Treat was not due to the thermal treatment, but to small localised porosity in the As-Dep sample due to process noise.

**Table 3**
Average density at 293 K for each sample set, the standard deviation of the measurements; the calculated density at 800 K and 1300 K; the density in percentage respect to the theoretical value ($\rho_R$).

| Samples Set | Avg. $\rho$ at 293 K [g/cm$^3$] | Standard Dev. | Calc. $\rho$ at 800 K [g/cm$^3$] | Calc. $\rho$ at 1300 K [g/cm$^3$] | $\rho_R$ [%] |
|---|---|---|---|---|---|
| Long As-Dep | 18.82 | 0.20 | 18.67 | 18.53 | 97.5 |
| Vert As-Dep | 19.1 | 0.05 | 18.94 | 18.77 | 98.9 |
| Long Heat-Treat | 19.17 | 0.03 | 19.01 | 18.83 | 99.3 |
| Vert Heat-Treat | 19.18 | 0.07 | 19.04 | 18.86 | 99.4 |

*3.6. Specific heat capacity*

**Fig. 11** reports the variation of the specific heat capacity at constant pressure over the range of temperatures from 300 K to 1300 K. In general, the curves present similar trends for both As-Dep and Heat-Treat samples. In particular, for all tested conditions, the heat capacity tended to increase with the temperature. An abrupt increment in heat capacity, occurring at around 873 K, was seen in both the As-Dep Long and Heat-Treat Long samples and measured as 25% and 14%, respectively. The sudden variation could have been caused by the recovery of crystal defects. The rearrangement of the atoms to eliminate point or line defects, as a thermally-activated process, draws energy without increasing the total temperature of the metal. The increment is less severe in the Heat-Treat samples because a possible recovery could have already occurred during the heat treatment. The differences between testing direction are not depended on the heat treatment, and can be mainly attributed to the anisotropy of the microstructure.

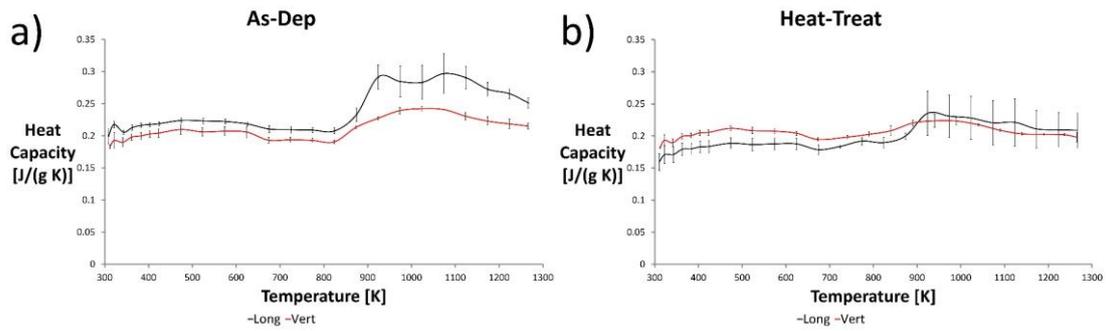

**Fig. 11**: Specific heat capacity curves against temperate of the tested samples in the As-Dep (a) and Heat-Treat condition (b) with the associated error bars.

*3.7.    Thermal conductivity*

Values of thermal conductivity (k) are reported in **Fig. 12**. It can be seen that k decreases by approx. 35% as the temperature increases from 300 K to 1300 K. The same abrupt changes seen in the heat capacity values are also found in k, as the latter is derived from the former. Overall, the trends reported resulted to be in accordance with other measurements of thermal conductivity for unalloyed tungsten [33,34]. The anisotropy in the microstructure and the application of the heat treatment had no considerable effect on the reduction of k over the temperature range.

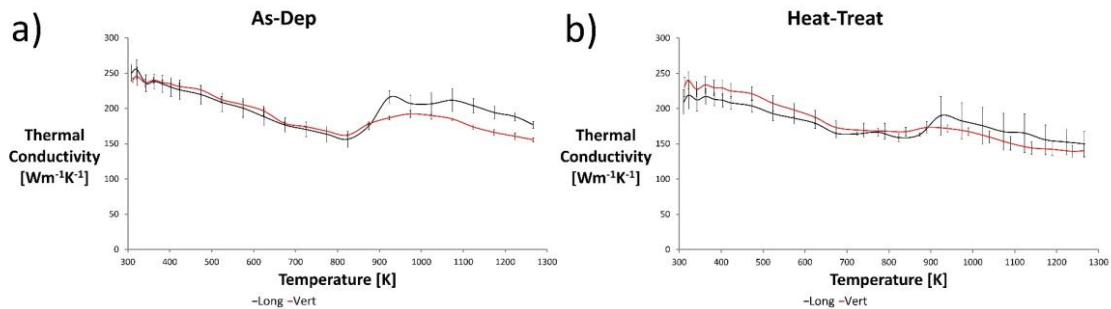

**Fig. 12**: Thermal conductivity curves against temperate of the tested samples in the As-Dep (a) and Heat-Treat condition (b) with the associated error bars.

Both thermal diffusivity and conductivity greatly depend on the density, a higher density yielding a higher thermal conductivity. In fact, it has been already reported that thermal conductivity can be influenced by volume defects [39,40]; it has also been reported that a high number of dislocation and/or grain boundaries can decrease the thermal conductivity [41]. This is because these features increase the interfacial area. In this study, surprisingly the As-Dep samples are characterised by a slightly lower density, but a higher thermal conductivity when compared to the Heat-Treat samples. Furthermore, the thermal conductivity of tungsten deposited by WAAM was found to be higher than that of other manufacturing methods [33,34,41,42], which can be explained by the higher purity, the higher density, and the lower number of grains boundaries seen in the structures presented in this research.

## 4    Conclusion

In this research, the microstructure and the thermal properties of WAAM-deposited tungsten have been characterised, and their relationships explained. The main results can be summarised as:

- The deposition strategy and the variation in thermal conductivity of tungsten promoted the evolution of a microstructure characterised by two specular arrays of columnar grains;

- The heat-treatment process neither had an effect on the microstructure nor on the thermal diffusivity and conductivity. However, it seems to have promoted, to some extent, stress-relaxation as can be deducted from the dilatometry data;

- The heat-treatment seemed to have also promoted recovery, which could be inferred from the comparison of the specific heat capacity of the as-deposited and the heat-treated samples;

- The thermal conductivity of tungsten components made by WAAM seems to show higher values compared to other manufacturing methods. This was attributed to WAAM specimens' higher purity, higher density, and lower number of grains boundaries.

This study demonstrated that the WAAM process can produce tungsten components with a suitable level of integrity to meet the requirements of the nuclear industry. Further characterisation of the effect of post-deposition heat treatments on residual stresses will be needed in order to understand the best production route for real industrial tungsten components. Additional metallographic tests are needed to understand the recovery of the structure deposited after heat treatment.

## Acknowledgement

The authors wish to acknowledge financial support from the AMAZE Project, which was co-funded by the European Commission in the 7th Framework Programme (contract FP7-2012-NMP-ICT-FoF-313781), by the European Space Agency and by the individual partner organisations.